\documentclass[a4paper,11pt]{article}
\usepackage{todonotes}

\usepackage[margin=0.8in,top=0.65in,bottom=0.65in]{geometry}
\usepackage{adjustbox}
\usepackage{algorithm}
\usepackage{algorithmic}
\usepackage{amsmath,amssymb,amsfonts,amsthm} 
\usepackage{array}
\usepackage{blkarray}
\usepackage{bm}
\usepackage{booktabs}
\usepackage{cite} 
\usepackage[numbers]{natbib}  
\usepackage{hyperref}
\hypersetup{
    colorlinks,
    linkcolor={red!50!black},
    citecolor={blue!50!black},
    urlcolor={blue!80!black}
}
\usepackage{cleveref}
\crefname{figure}{figure}{figures}
\usepackage{diagbox}
\usepackage{dsfont}
\usepackage{enumerate}
\usepackage[inline]{enumitem}
\usepackage{graphicx}
\usepackage{mathtools}
\usepackage{multirow}
\usepackage{nth}
\usepackage{siunitx}
\usepackage{soul}
\usepackage{stmaryrd}
\usepackage{subcaption}
\usepackage{textcomp}
\usepackage{xcolor}
\usepackage{parskip}
\usepackage{url}
\usepackage{tikz}
\usepackage{tikzsymbols}
\usepackage{pgfplots}
\pgfplotsset{compat=newest} 
\usetikzlibrary{arrows,shapes}
\usetikzlibrary{decorations.markings}
\usetikzlibrary{decorations.pathmorphing}
\usetikzlibrary{arrows.meta}
\usetikzlibrary{shadows,arrows,positioning,shapes.geometric}
\tikzset{cross/.style={cross out, draw=black, inner sep=0pt, outer sep=0pt},cross/.default={1pt}}
\usepackage{tikzsymbols}
\usepackage{pgfplots}
\pgfplotsset{compat=newest} 
\usetikzlibrary{plotmarks}
\usetikzlibrary{arrows.meta}
\usepgfplotslibrary{patchplots}
\usepackage{grffile}
\usepackage{pdflscape}
\usepackage{fontawesome}
\pgfplotsset{every axis/.append style={
                    label style={font=\scriptsize},
                    tick label style={font=\scriptsize},
                    legend style={font=\scriptsize}
                    }}
\pgfplotsset{compat=newest}
\pgfplotsset{plot coordinates/math parser=false}
\pgfplotsset{grid style={dotted,gray}}
\pgfplotsset{
compat=1.11,
legend image code/.code={
\draw[mark repeat=2,mark phase=2]
plot coordinates {
(0cm,0cm)
(0.15cm,0cm)        
(0.3cm,0cm)         
};%
}
}
\usepgfplotslibrary{groupplots}
\newlength\figureheight
\newlength\figurewidth
\pgfdeclarelayer{background}
\pgfdeclarelayer{foreground}
\pgfsetlayers{background,main,foreground}

\newtheorem{remark}{Remark}

\renewcommand{\vec}[1]{\boldsymbol{#1}}

\newcommand{\flowmap}{\vec{\Phi}}
\newcommand{\submap}{\flowmap}
\newcommand{\scalar}{f}

\newcommand{\R}{\ensuremath{\mathbb{R}}} 

\newcommand{\dx}{\ensuremath{{\mathrm d}x}} 
\newcommand{\dv}{\ensuremath{{\mathrm d}v}} 

\newcommand{\Nremap}{N_\text{remap}}
\newcommand{\mapcmm}{\vec{\chi}}

\newcommand{\Nsample}{N_{f}}

\newcommand{\energy}{\mathcal{E}}

\newcommand{\Epot}{\energy_\mathrm{pot}}                 
\newcommand{\Epond}{E_\mathrm{pond}}

\usepackage{xcolor}

\usepackage[acronym,automake,nonumberlist]{glossaries}
\newcommand{\VP}{{VP}\xspace}
\newcommand{\CMM}{{CMM}\xspace}
\newcommand{\NuFi}{{NuFi}\xspace}

\makeglossaries

\begin{document}


\title{Revisiting kinetic electrostatic electron non-linear (KEEN) waves in the presence of dynamical ions}
 \date{\today}

\renewcommand{\and}{\end{tabular}\hskip 0.1em\begin{tabular}[t]{c}}
\author{
     R.-Paul Wilhelm\thanks{Centre for Mathematical Plasma Astrophysics, Department of Mathematics, KU Leuven, B-3001 Leuven, Belgium} 
    \and Philipp Krah\thanks{IRFM - CEA Cadarache,
    13108 Saint-Paul-lez-Durance, France }
    \and Kai Schneider\thanks{Aix-Marseille Université, I2M, CNRS, UMR 7373, 3 place Victor Hugo, 13331 Marseille cedex 3, France.}
    \and Fabio Bacchini\footnotemark[1] \thanks{Royal Belgian Institute for Space Aeronomy, Solar-Terrestrial Centre of Excellence, B-1180 Uccle, Belgium}
    \and Virginie Grandgirard\footnotemark[2]
}

%
%

\maketitle


\begin{abstract}
We revisit the kinetic electrostatic electron nonlinear (KEEN) waves studied by Afeyan et. al in 2014 using a hybrid flow-mapping strategy that combines the characteristic mapping method (CMM) with numerical flow iteration (NuFi).
The study extends the classical setup to dynamical ions, comparing their impact on the long-time KEEN dynamics with the static-ion case.
To this end, we extend the CMM-NuFI framework with a multi-map strategy, assigning one map to the ion and one to the electron characteristic flow.
The resulting problem exhibits a wide separation of spatial and temporal scales, driven by the fine structures generated by the ponderomotive force and by the large ion-to-electron mass ratio, which renders it computationally prohibitive for conventional grid-based methods.
Our multi-map CMM-NuFI method efficiently resolves these disparate scales, enabling long-time simulations of KEEN dynamics with fully dynamical ions.
\end{abstract}
 
    
\providecommand{\placeholderfig}[2]{%
  \fbox{\begin{minipage}[c][#1][c]{0.92\linewidth}
    \centering\small #2
  \end{minipage}}%
}
\providecommand{\mapgrids}[1]{\mathcal{G}_{\mapcmm,#1}}
\providecommand{\Nmapgrids}[1]{N_{\mapcmm,#1}}
\providecommand{\Nremaps}[1]{N_{\mathrm{remap},#1}}
\section{Introduction}
\label{sec:intro}

Kinetic electrostatic electron non-linear (KEEN) waves are self-organized, non-stationary electron phase-space structures that arise in the driven, ion-static Vlasov--Poisson (\VP) system without belonging to any linear dispersion branch or Bernstein--Greene--Kruskal (BGK) equilibrium~\cite{AfeyanWonSavchenkoJohnstonGhizzoBertrand2003}. They are excited by the ponderomotive force of two crossing laser beams, which couples into the electron distribution $f$ through an external drive field,
\begin{equation}
\label{eq:vp-static}
\partial_t f + v\partial_x f + (E-\Epond)\partial_v f = 0,\qquad
\partial_x E = \rho = 1-\int_\R f\,\dv,
\end{equation}
where $x$ and $v$ are the spatial and velocity coordinates of the 1D1V \VP{} system, $t$ is time, $E$ is the self-consistent electric field, and $\rho$ is the charge density. The laser-induced ponderomotive force is represented by
\begin{equation}
\label{eq:pond-intro}
\Epond(x,t) = a_\mathrm{Dr}k_\mathrm{Dr}a(t)\sin(k_\mathrm{Dr}x-\omega_\mathrm{Dr}t),
\end{equation}
with a smooth adiabatic switch $a(t)$ that turns the drive on and off over a finite duration $T_\mathrm{Dr}$. Unlike electron-acoustic waves or BGK modes, which require a distribution function cobbled to sit exactly on a resonance curve, KEEN waves form anywhere in the $(\omega,k)$ plane once the drive is strong and of long enough duration, and persist as phase-locked, multi-harmonic structures long after it is switched off~\cite{AfeyanCasasCrouseillesDodhyFaouMehrenbergerSonnendruecker2014}. In equation~\eqref{eq:vp-static}, and in essentially all KEEN wave studies to date, ions only enter as a fixed, uniform neutralizing background.

Numerically, KEEN waves are notoriously demanding. 
Their self-organization proceeds through shedding, merging, and re-trapping of many small phase-space vorticlets around the driven phase velocity, producing filaments whose width scales with the square root of the local field amplitude~\cite{AfeyanCasasCrouseillesDodhyFaouMehrenbergerSonnendruecker2014}. 
Resolving these structures over the long times ($t\sim10^3$ inverse electron plasma frequencies, $\omega_\mathrm{pe}^{-1}$) needed for a KEEN wave to self-organize requires prohibitively fine velocity resolution if a uniform grid is used everywhere. 
This is addressed in \cite{AfeyanCasasCrouseillesDodhyFaouMehrenbergerSonnendruecker2014} with a two-grid, non-uniform, conservative cubic spline semi-Lagrangian scheme and sixth-order time-splitting, which still need grids up to $8192\times16384$ points (canonical drive case) and report that the low-order Fourier modes of $\rho$ have not fully converged even at these resolutions. 
Reaching this fidelity costs on the order of $10^4$--$10^5$ CPU-hours per run, which makes extending the model, e.g.\ by including ion dynamics, prohibitively expensive with such grid-based solvers. Furthermore, the non-uniform grid approach introduces subshocks in the transition regions between the different cell sizes, which is visible in the distribution function \citep[Fig. 2]{AfeyanCasasCrouseillesDodhyFaouMehrenbergerSonnendruecker2014}.

An efficient alternative is offered by \emph{flow-map} methods, which evolve the backward characteristic map $\flowmap_0^t$ of the \VP{} system instead of $f$ itself, using $f(x,v,t)=f_0(\flowmap_0^t(x,v))$, where $f_0=f(\cdot,\cdot,0)$ is the initial distribution, together with the semigroup property $\flowmap_0^t=\submap_0^\tau\circ\submap_\tau^{2\tau}\circ\cdots\circ\submap_{(n-1)\tau}^{t}$. The Numerical Flow Iteration (\NuFi) advances this composition iteratively, backward in time, without ever storing $f$ on a grid, at the cost of a per-time-step iteration count that grows with the simulated time~\cite{KirchhartWilhelm2024}. The Characteristic Mapping Method (\CMM) instead stores the composition of a handful of low-order polynomial submaps~\cite{MercierYinNave2019,NaveRosalesSeibold2010,YinMercierYadavSchneiderNave2021,KrahYinBergmannNaveSchneider2024}, each spanning many elementary time steps at once. The hybrid CMM-\NuFi{} method~\cite{KrahLinBacchiniNaveGrandgirardSchneider2026} combines both, inheriting \NuFi's favorable conservation and resolution properties while significantly reducing its quadratic cost growth in time, and has already reproduced the canonical, ion-static KEEN wave of~\cite{AfeyanCasasCrouseillesDodhyFaouMehrenbergerSonnendruecker2014} on sample grids more than two orders of magnitude coarser than the reference solution (see \citep[section 3.3]{KrahLinBacchiniNaveGrandgirardSchneider2026}).

\textbf{Contribution.} We extend the canonical KEEN wave setup to a two-species \VP{} system with dynamical ions (\cref{sec:dynion}), and generalize the CMM-\NuFi{} algorithm to multiple species by maintaining one independent set of characteristic submaps per species (\cref{sec:multimap}), building on multi-species \NuFi~\cite{Wilhelm_2025,WilhelmTorrilhon2026}. \Cref{sec:results} compares the resulting canonical KEEN wave with dynamical and static ions in terms of the electron density, the $\rho$-harmonics, and energy evolution.

\section{KEEN waves with dynamical ions}
\label{sec:dynion}

We extend \eqref{eq:vp-static} to a two-species \VP{} system for electrons ($f_e$, charge $q_e$, mass $m_e$) and a single ion species ($f_i$, charge $q_i$, mass $m_i$). It is convenient to collect $q_e,m_e,q_i,m_i$ into the charge number $Z_i\equiv-q_i/q_e$ and the electron-to-ion mass ratio $\mu\equiv m_e/m_i$. Normalizing space to the electron Debye length, time to $\omega_\mathrm{pe}^{-1}$, velocities to the electron thermal speed $v_{\text{th},e}$, and charge-to-mass ratios so that $q_e/m_e=1$, the two species obey
\begin{align}
\label{eq:vlasov-electron}
\partial_t f_e + v\partial_x f_e + \frac{q_e}{m_e}(E-\Epond)\partial_v f_e &= 0,\\
\label{eq:vlasov-ion}
\partial_t f_i + v\partial_x f_i + \frac{q_i}{m_i}(E-\Epond)\,\partial_v f_i &= 0,
\end{align}
coupled through the electric field via the total charge density,
\begin{equation}
\label{eq:poisson-2species}
\partial_x E = Z_i n_i - n_e,\qquad
n_s(x,t)=\int_\R f_s\,\dv,\quad s\in\{e,i\}.
\end{equation}
The ponderomotive drive $\Epond$ of equation~\eqref{eq:pond-intro} enters both species through the same charge-to-mass ratio that multiplies the self-consistent field, $q_e/m_e=1$ for electrons and $q_i/m_i=-\mu Z_i$ for ions: since $q_i/m_i$ is small whenever $\mu Z_i\ll1$, the ponderomotive term is correspondingly weaker on ions. The sign of $q_i/m_i$ is opposite to $q_e/m_e$ because ions and electrons carry opposite charge, and its small magnitude reflects the much smaller acceleration produced by the same force on the heavier ions, setting $\mu\to0$ (equivalently $q_i/m_i\to0$) recovers the ion-static system \eqref{eq:vp-static} with $n_i\equiv1$, $Z_i=1$. Because $\mu Z_i\ll1$ for the hydrogen ions considered here ($\mu=1/1836$), the direct action of $\Epond$ on the ions is expected to be negligible compared to its effect on the electrons, so ions still respond to the drive predominantly indirectly, through the self-consistent field $E$.

\begin{remark}
Physically, $\Epond$ represents the ponderomotive force of the crossing laser beams, an external forcing rather than a self-consistent electrostatic field, so writing it as a potential contribution to $E$ is a modeling simplification. For the hydrogen ions used here ($Z_i=1$), and equivalently in the ion-static limit, this simplification is exact. For heavier or more highly charged ion species, however, the drive would need to be applied as a genuine external force distinct from $E$, which would also require revisiting how the drive enters the CMM-\NuFi{} scheme.
\end{remark}

We use the same canonical drive and adiabatic switch as the ion-static reference case~\cite{AfeyanCasasCrouseillesDodhyFaouMehrenbergerSonnendruecker2014},
\begin{equation}
\label{eq:drive-envelope}
a(t)=\frac{g(t)-g(t_0)}{1-g(t_0)},\qquad
g(t)=\tfrac12\left(\tanh\tfrac{t-t_L}{t_{wL}}-\tanh\tfrac{t-t_R}{t_{wR}}\right),
\end{equation}
with $t_0=0$, $t_L=69$, $t_{wL}=t_{wR}=20$, $t_R=207+T_\mathrm{Dr}$, drive wavenumber $k_\mathrm{Dr}=0.26$, frequency $\omega_\mathrm{Dr}=0.37$, amplitude $a_\mathrm{Dr}=0.2$, and duration $T_\mathrm{Dr}=100$ (the \emph{canonical} drive). Both species start from Maxwellian distributions at rest (recall that $v_{\mathrm{th},e}=1$ in our units),
\begin{equation}
\label{eq:initial-condition}
f_{e,0}(x,v)=\frac{1}{\sqrt{2\pi}}e^{-v^2/2},\qquad
f_{i,0}(x,v)=\frac{1}{\sqrt{2\pi}\,v_{\text{th},i}}e^{-v^2/2v_{\text{th},i}^2},\qquad
v_{\text{th},i}=\sqrt{\mu/Z_i},
\end{equation}
assuming equal electron and ion temperatures, on $(x,v)\in[0,2\pi/k_\mathrm{Dr})\times[-6,6]$ for electrons and $(x,v)\in[0,2\pi/k_\mathrm{Dr})\times[-6v_{\text{th},i},6v_{\text{th},i}]$ for ions. We consider hydrogen ions, $Z_i=1$, $\mu=1/1836$, so that $v_{\text{th},i}\approx0.0233$: the ion phase space is thus $\sim\!43$ times narrower in velocity than the electron one, reflecting the physical scale separation that the numerical method must resolve simultaneously for both species.

\section{Multi-map CMM-\NuFi{} for multiple species}
\label{sec:multimap}
In the following, we describe the generalization of the CMM-\NuFi{} method to multiple species.

\subsection*{Multi-species \NuFi}
For a single species $s$, the solution of equations~\eqref{eq:vlasov-electron}--\eqref{eq:vlasov-ion} can be written, following~\cite{Wilhelm_2025,WilhelmTorrilhon2026}, as
\begin{equation}
\label{eq:multispecies-flowmap}
f_s(t,x,v) = f_{s,0}\big(\flowmap^{t,s}_{0}(x,v)\big),
\end{equation}
where the backward flow map $\flowmap^{t,s}_{0}$ of species $s$ is defined through the characteristic ODE
\begin{equation}
\label{eq:multispecies-ode}
\frac{\mathrm d}{\mathrm d\sigma}\hat x_s(\sigma) = -\hat v_s(\sigma),\qquad
\frac{\mathrm d}{\mathrm d\sigma}\hat v_s(\sigma) = -\frac{q_s}{m_s}\big(E(\sigma,\hat x_s(\sigma))-\Epond(\sigma,\hat x_s(\sigma))\big),
\end{equation}
with $\hat x_s(t)=x$, $\hat v_s(t)=v$, integrated backward from $\sigma=t$ to $\sigma=0$. \NuFi{} approximates equation~\eqref{eq:multispecies-ode} with a St\"ormer--Verlet step~\cite{KirchhartWilhelm2024}: starting from $\hat x^h_{s,i}=x$, $\hat v^h_{s,i}=v$ at time step $t_i$, compute
\begin{align}
\label{eq:multispecies-sv1}
\hat v^h_{s,i-1/2} &= \hat v^h_{s,i} - \frac{\tau}{2}\frac{q_s}{m_s}\big(E(t_i,\hat x^h_{s,i})-\Epond(t_i,\hat x^h_{s,i})\big),\\
\label{eq:multispecies-sv2}
\hat x^h_{s,i-1} &= \hat x^h_{s,i} - \tau\,\hat v^h_{s,i-1/2},\\
\label{eq:multispecies-sv3}
\hat v^h_{s,i-1} &= \hat v^h_{s,i-1/2} - \frac{\tau}{2}\frac{q_s}{m_s}\big(E(t_{i-1},\hat x^h_{s,i-1})-\Epond(t_{i-1},\hat x^h_{s,i-1})\big),
\end{align}
for $i=N_s,N_s-1,\dots,1$, so that $f_s(t,x,v)=f_{s,0}(\hat x^h_{s,0},\hat v^h_{s,0})+\mathcal O(\tau^2)$. Since \eqref{eq:multispecies-ode}--\eqref{eq:multispecies-sv3} depend on the species only through the charge-to-mass ratio $q_s/m_s$ ($q_e/m_e=1$ for electrons, $q_i/m_i=-\mu Z_i$ for ions), the same iteration applies unchanged to every species $s\in\{e,i\}$. Both species nonetheless use the same, shared electric field history $[E^{(n)}]_n$: it is the coupling \eqref{eq:poisson-2species} through the densities, not the flow map itself, that links electrons and ions. Concretely, at every time step the densities of {both} species are accumulated by tracing their (independent) backward maps to the initial condition, before the shared Poisson equation \eqref{eq:poisson-2species} is solved in Fourier space for the field that both species will use in the next step.

\subsection*{Generalization of CMM-\NuFi{} to multiple maps}
\Cref{alg:multi-CMM-NuFi} generalizes the hybrid CMM-\NuFi{} time step of~\cite{KrahLinBacchiniNaveGrandgirardSchneider2026} to a set of species $\mathcal S$ by keeping one independent list of submaps $[\mapcmm_s^{(k)}]_{k=1,\dots,M_s}$, one \NuFi{} counter $N_s$, and one coarse map grid $\mapgrids{s}$ per species $s\in\mathcal S$; only the sample-grid loop that accumulates the densities $n_s$, and the resulting Poisson solve, are shared between species. The subroutines \texttt{NuFi} and \texttt{MapsCompose} are unchanged from the single-species algorithm in~\cite{KrahLinBacchiniNaveGrandgirardSchneider2026}, {i.e., using cubic spline interpolation for the maps and 2nd-order symplectic time integration}. In each time step, each species performs one \NuFi{} step of its own iterative chain and, once $N_s$ reaches the remapping frequency $\Nremap$, that chain collapses into a new submap $\mapcmm_s^{(M_s+1)}$, exactly as in the single-species algorithm. Species can in principle use different remapping frequencies $\Nremaps{s}$ and map-grid resolutions $\Nmapgrids{s}$. For example, the slowly evolving ions could use a coarser, less frequently updated map. For \cref{sec:results}, however, we use identical parameters for both species. Because ions move comparatively little on the electron time scale $\Delta t=\Nremap\tau$, their submaps remain well approximated on the same coarse grid as the electron submaps.

\begin{algorithm}[!t]
\small
\begin{algorithmic}[1]
\renewcommand{\COMMENT}[2][.42\linewidth]{\leavevmode\hfill\makebox[#1][l]{//~#2}}
\renewcommand{\algorithmicrequire}{\textbf{Input:}}
\renewcommand{\algorithmicensure}{\textbf{Output:}}
\REQUIRE{Species $\mathcal S$ with charge-to-mass ratios $q_s/m_s$, charges $Z_s$, submaps $[\mapcmm_s^{(k)}]_{k\le M_s}$, counters $N_s$, coarse grids $\mapgrids{s}$ ($s\in\mathcal S$), field history $[E^{(n)}]_n$, remapping frequency $\Nremap$, time step $\tau$, sample grid resolution $\Nsample$.}
\ENSURE{Updated submaps $[\mapcmm_s^{(k)}]_{k,s}$ and field history $[E^{(n)}]_n$.}
\FOR{$s\in\mathcal S$}
  \STATE{$N_s\leftarrow N_s+1$}
\ENDFOR
\FOR{$j=0,\dots,\Nsample-1$}
  \STATE{$\rho_j\leftarrow 0$}
  \COMMENT{accumulate $Z_in_i-n_e$ at $x_j$}
  \FOR{$s\in\mathcal S$}
    \FOR{$k=0,\dots,\Nsample-1$}
      \STATE{$(x,v)\leftarrow$ sample grid point of species $s$}
      \STATE{$(x,v)\leftarrow \texttt{NuFi}((x,v),[\tfrac{q_s}{m_s}(E^{(n)}-\Epond)]_n,\tau,N_s,\texttt{false})$}
      \COMMENT{scaled by $q_s/m_s$ for each species}
      \STATE{$(x^*,v^*)\leftarrow \texttt{MapsCompose}([\mapcmm_s^{(k)}]_k,(x,v))$}
      \STATE{$\rho_j\leftarrow \rho_j + Z_s\, \scalar_{s,0}(x^*,v^*)\,\Delta v_s$}
      \COMMENT{$Z_e:=-1$}
    \ENDFOR
  \ENDFOR
\ENDFOR
\FOR{$s\in\mathcal S$}
  \IF{$N_s\bmod \Nremap = 0$}
    \STATE{$M_s\leftarrow M_s+1$}
    \FOR{$(x,v)$ on coarse grid $\mapgrids{s}$}
      \STATE{$\mapcmm_s^{(M_s)}(x,v)\leftarrow \texttt{NuFi}((x,v),[\tfrac{q_s}{m_s}(E^{(n)}-\Epond)]_n,\tau,N_s,\texttt{true})$}
    \ENDFOR
    \STATE{$N_s\leftarrow 0$}
  \ENDIF
\ENDFOR
\STATE{Solve Poisson \eqref{eq:poisson-2species} for $E^{(n)}$}
\end{algorithmic}
\caption{Multi-species CMM-\NuFi{} time step, generalizing the single-species algorithm of~\cite{KrahLinBacchiniNaveGrandgirardSchneider2026} to a species set $\mathcal S$ (here $\mathcal S=\{e,i\}$, $M_s=|\mathcal S|=2$ maps).}
\label{alg:multi-CMM-NuFi}
\end{algorithm}

\section{Results}
\label{sec:results}

We simulate the canonical KEEN wave of \cref{sec:dynion} up to $t=2000$ with the multi-species CMM-\NuFi{} method of 
\cref{sec:multimap}, using time step $\tau=0.05$, remapping frequency $\Nremap=200$ (one new submap every 
$\Delta t=20$), and a sample grid of $N_x\times N_v=1024\times 512$ points in the two-dimensional $(x,v)$ phase 
space of each species, allowing up to 5 levels of refinement for the velocity quadrature rule with a relative 
tolerance of $10^{-5}$ for refinement and $10^{-8}$ for stopping. 
The maps are stored on coarse grids $\mapgrids{e},\mapgrids{i}$ of $64 \times 256$. 
This yields $M_s=|\mathcal S|=2$ independently maintained CMM maps over the whole run, one for electrons and one for ions. We compare this dynamical-ion run ($\mu=1/1836$, $Z_i=1$) against the ion-static reference obtained by setting $\mu=0$ in \eqref{eq:vlasov-ion} (equivalently $n_i\equiv1$), keeping the drive, grids, and time-stepping otherwise identical.

\subsubsection*{Electron density}
The zoom in \cref{fig:density_comparison} compares the deviation of the electron distribution function $f_e(t,x,v)$ from the initial Maxwellian for the ion-static and 
ion-dynamic runs. Since the direct action of $\Epond$ on ions is suppressed by the mass ratio $\mu Z_i\ll1$, the ion dynamics are delayed and thus the electron distribution functions are expected to agree closely while the drive is on and shortly after. Any deviation, driven by the slowly evolving ions coupling back into $E$ through equation~\eqref{eq:poisson-2species}, should only become visible on the longer, ion time scale $t\gtrsim500\,\omega_{p,e}^{-1}$.
This can be observed when comparing the electron distribution functions at $t=300\,\omega_{p,e}^{-1}$ for the stationary ion case (see \cref{subfig:canonical_with_static_ions_f_minus_f0_t_300}), 
and for the mobile ion case (see \cref{subfig:canonical_with_static_ions_f_minus_f0_t_300}), which are largely similar. However, at $t=1000\,\omega_{p,e}^{-1}$ we already observe clear dynamical differences (see \cref{subfig:canonical_f_minus_f0_ion_1000} for the stationary and \cref{subfig:canonical_with_mobile_ions_f_minus_f0_t_1000} for the mobile ions case). Not only is the position of the main vortex different, but we also observe more small-scale phase-space vortices for larger $v$ in the mobile ion case, suggesting that more particle traps develop when including ion dynamics.

\begin{figure}
\centering
\begin{subfigure}{0.49\textwidth}
\caption{\label{subfig:canonical_with_static_ions_f_minus_f0_t_300}
$(f_e-f_{e,0})(300\,\omega_{p,e}^{-1})$, case: static ions}
\includegraphics[width=0.99\textwidth]{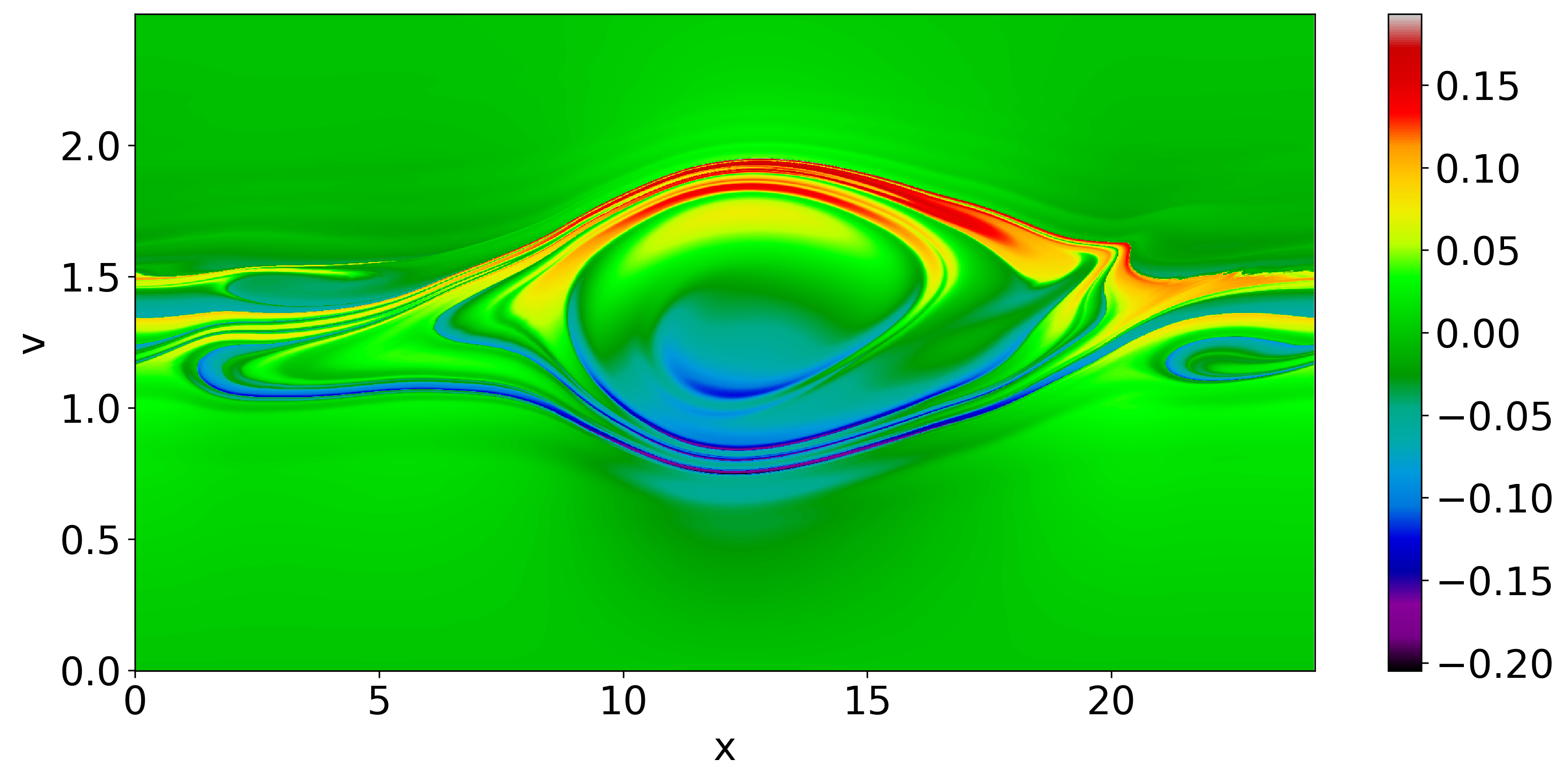}    
\end{subfigure}
\begin{subfigure}{0.49\textwidth}
\caption{\label{subfig:canonical_mobile_ions_f_minus_f0_300}
$(f_e-f_{e,0})(300\,\omega_{p,e}^{-1})$, case: dynamic ions}
\includegraphics[width=0.99\textwidth]{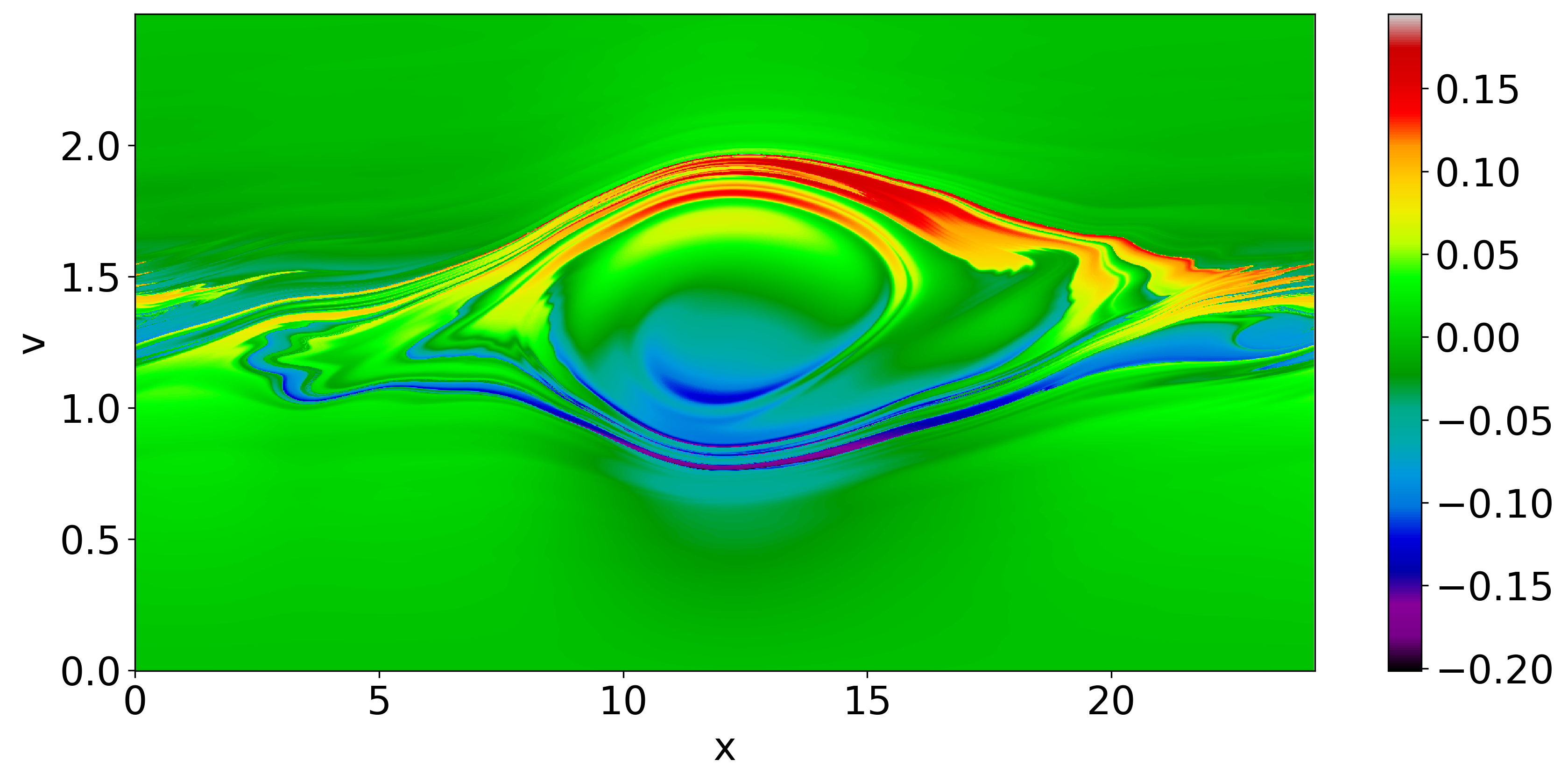}   
\end{subfigure}
\begin{subfigure}{0.49\textwidth}
\caption{\label{subfig:canonical_f_minus_f0_t_1000}
$(f_e-f_{e,0})(1000\,\omega_{p,e}^{-1})$, case: static ions }
\includegraphics[width=0.99\textwidth]{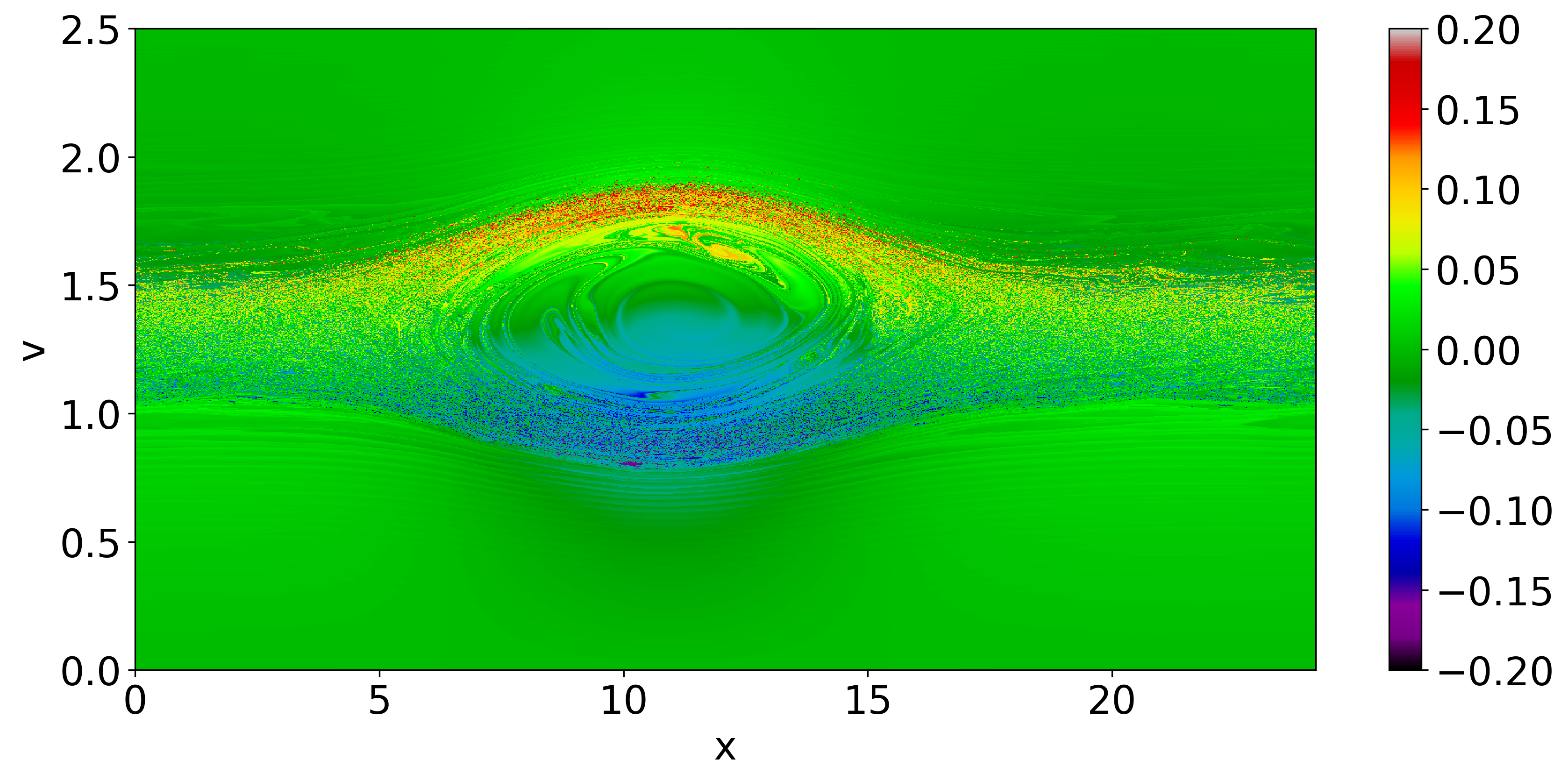}    
\end{subfigure}
\begin{subfigure}{0.49\textwidth}
\caption{\label{subfig:canonical_with_mobile_ions_f_minus_f0_t_1000}
$(f_e-f_{e,0})(1000\,\omega_{p,e}^{-1})$, case: dynamic ions}
\includegraphics[width=0.99\textwidth]{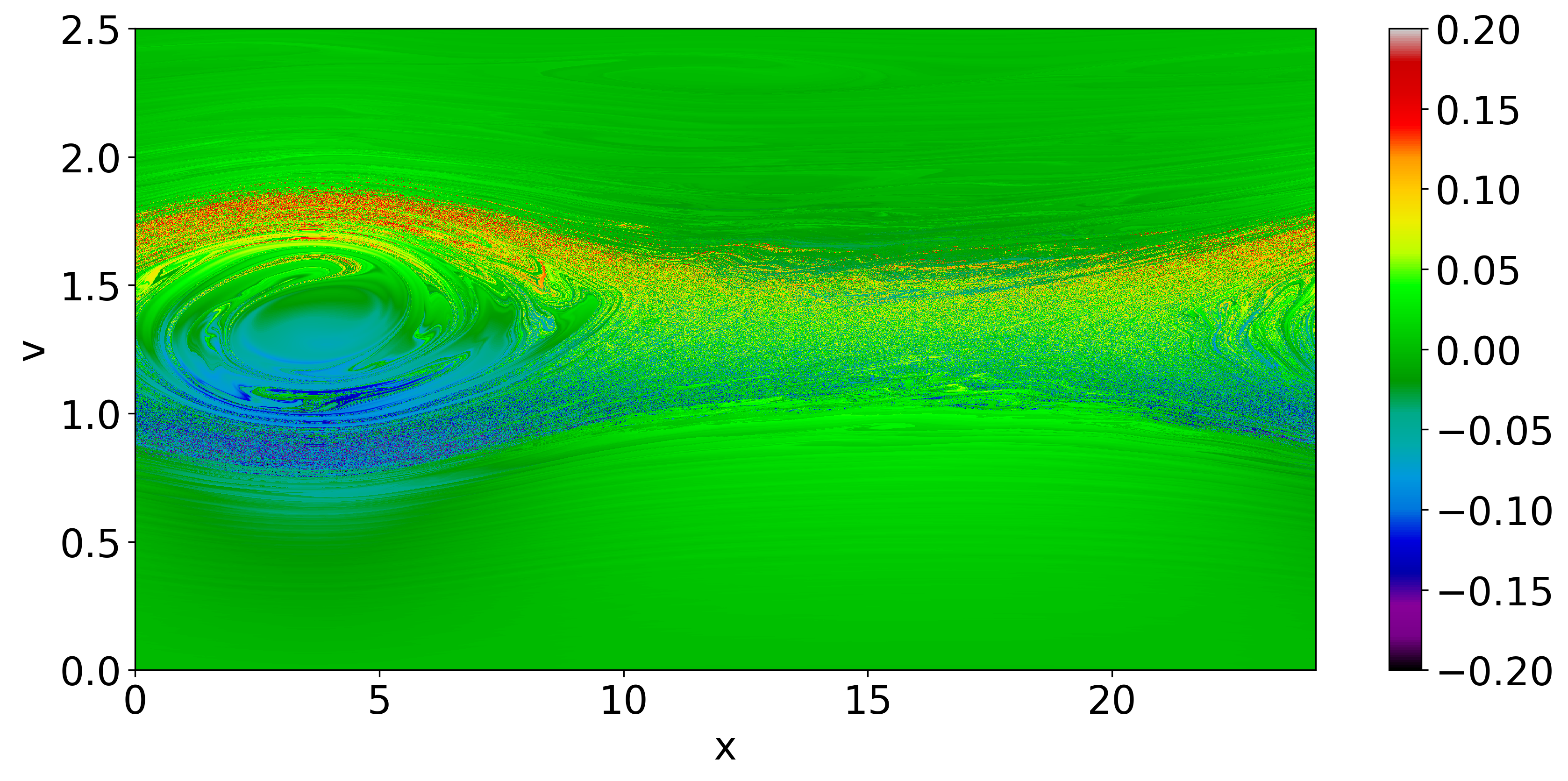}    
\end{subfigure}
\caption{Electron distribution function $f_e-f_{e,0}$ at $t=300,1000 \omega_{p,e}^{-1}$ for the canonical KEEN wave: ion-static reference (a,c) vs.\ ion-dynamic run with $\mu=1/1836$ (b,d), same drive and grid resolution in both cases.}
\label{fig:density_comparison}
\end{figure}

\subsubsection*{Ion distribution function}
\Cref{fig:ion_distribution} shows the ion distribution function $f_i(x,v,t)$ obtained from the ion submap $\mapcmm_i$ at time $t=1000\,\omega_{p,e}^{-1},2000\,\omega_{p,e}^{-1}$. Because $|q_i/m_i|=\mu Z_i\ll1$, ions are only weakly accelerated over the course of the simulation, so $f_i$ is expected to retain a nearly Maxwellian core with a small-amplitude, phase-locked perturbation.

\begin{figure}
\centering
\begin{subfigure}{0.49\textwidth}
\caption{\label{subfig:canonical_f_minus_f0_ion_1000}
$(f_i-f_{i,0})(1000\,\omega_{p,e}^{-1})$}
\includegraphics[width=0.99\textwidth]{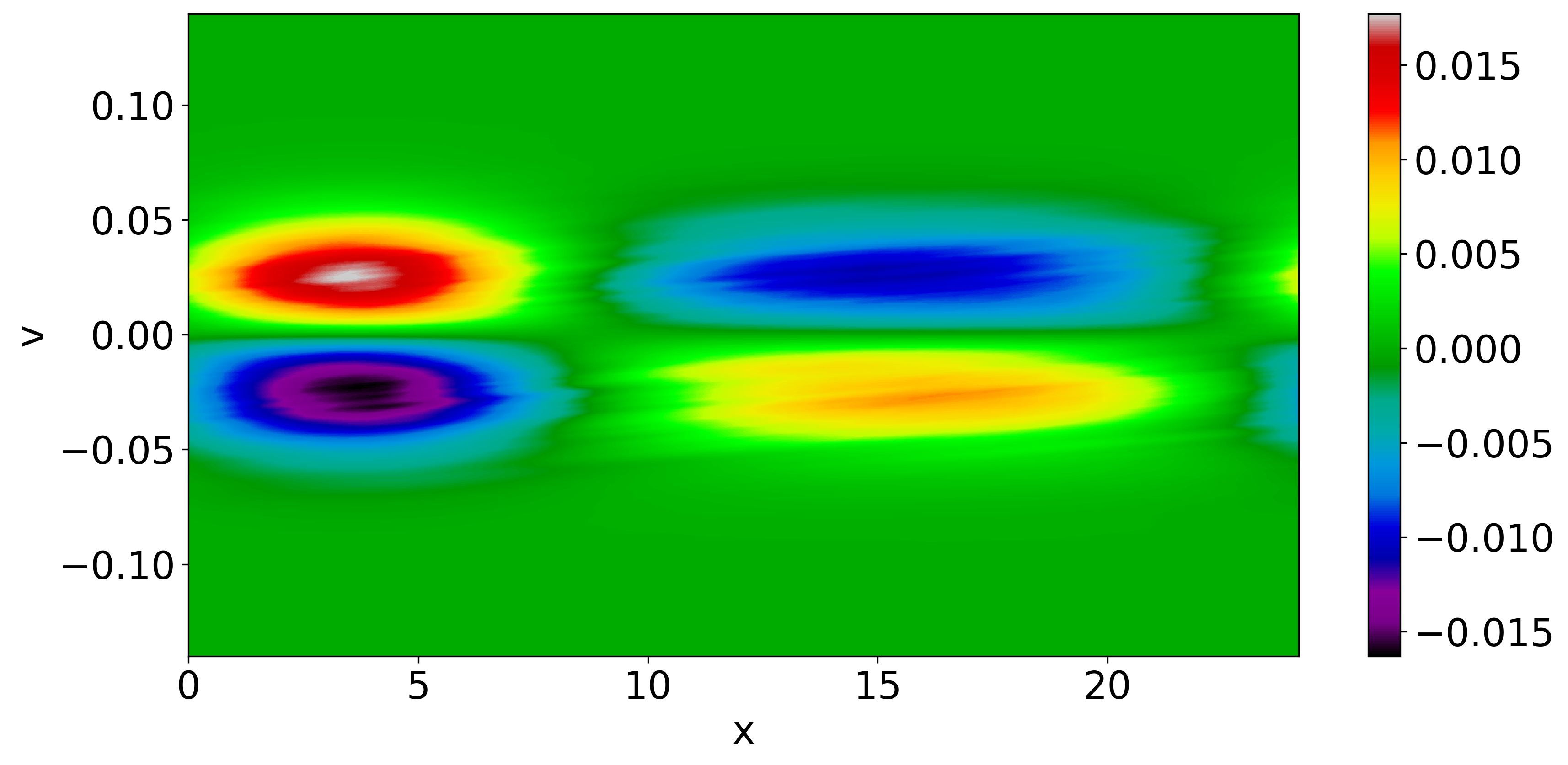}    
\end{subfigure}
\begin{subfigure}{0.49\textwidth}
\caption{\label{subfig:canonical_f_minus_f0_ion_2000}
$(f_i-f_{i,0})(2000\,\omega_{p,e}^{-1})$}
\includegraphics[width=0.99\textwidth]{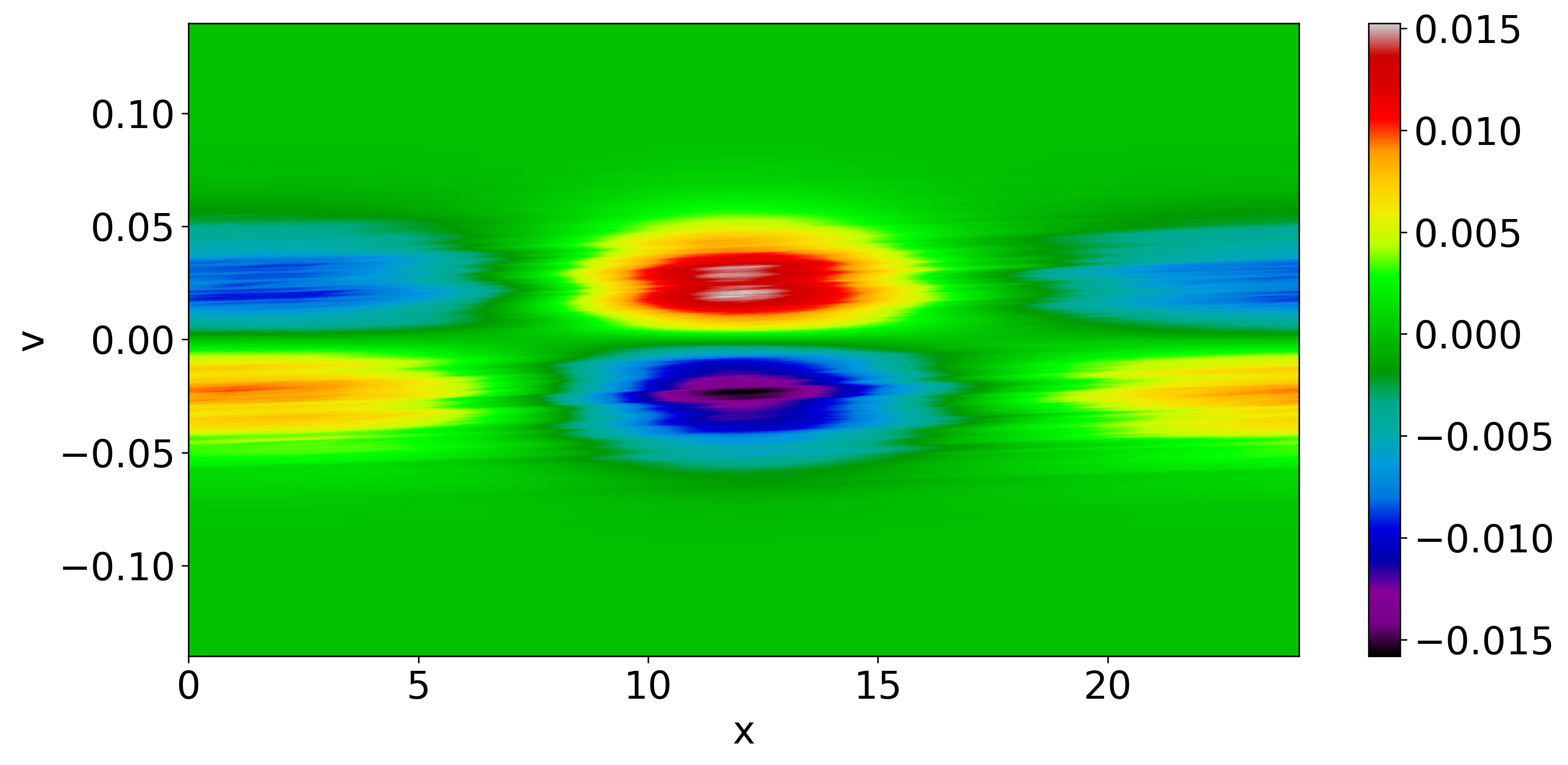}    
\end{subfigure}
\caption{Deviation of the ion distribution function from the initial condition $f_i - f_{i,0}$ at times a) $t=1000 \omega_{p,e}^{-1}$ and b) $t=2000\omega_{p,e}^{-1}$ in the ion-dynamic canonical run.}
\label{fig:ion_distribution}
\end{figure}

\subsubsection*{$\rho$-harmonics}
Following \cite{AfeyanCasasCrouseillesDodhyFaouMehrenbergerSonnendruecker2014}, we track the amplitudes of the five largest spatial Fourier modes ($k=1,\dots,5$) of the electron density $n_e(x,t)$, referred to as \emph{$\rho$-harmonics}. \Cref{fig:rho_harmonics} compares these harmonics between the ion-static and ion-dynamic runs. Agreement during the drive phase confirms that the fast electron dynamics that seed the KEEN wave are unaffected by including ion motion. At the same time, a slow relative drift of the harmonic amplitudes at late times quantifies the cumulative effect of the ion response on the sustained KEEN structure.

\begin{figure}
\centering
\setlength{\figurewidth}{0.9\linewidth}
\setlength{\figureheight}{0.4\linewidth}
\input{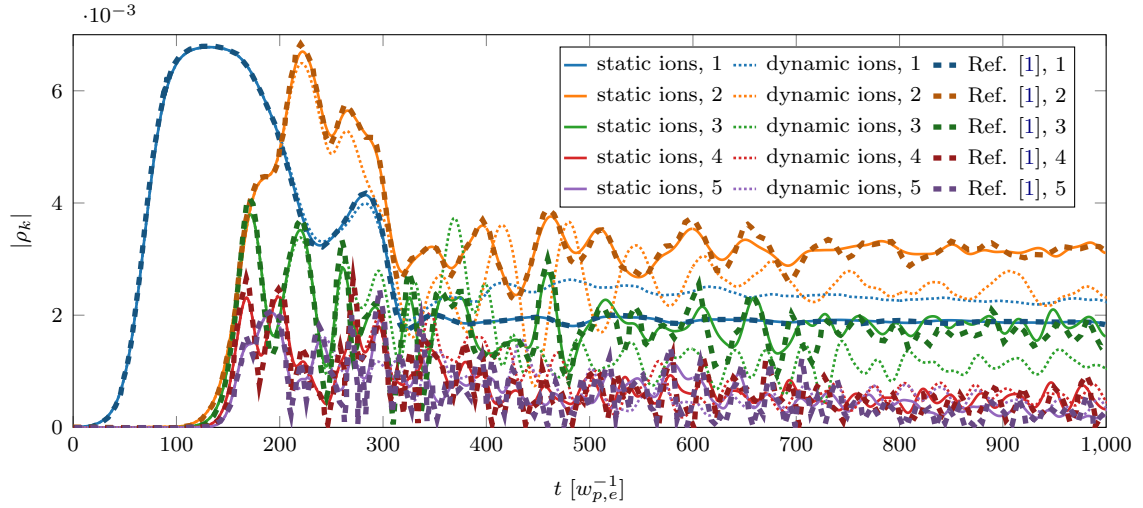}
\caption{Time evolution of the five largest $\rho$-harmonics of $n_e(x,t)$ for the ion-static and ion-dynamic canonical runs.}
\label{fig:rho_harmonics}
\end{figure}

\subsubsection*{Energy evolution}
We monitor the kinetic and potential energy, normalized consistently with equations~\eqref{eq:vlasov-electron}--\eqref{eq:vlasov-ion},
\begin{equation}
\label{eq:energies}
\energy_{\mathrm{kin},e}(t) = \frac12\iint v^2f_e\,\dx\dv,\quad
\energy_{\mathrm{kin},i}(t) = \frac{1}{2\mu}\iint v^2f_i\,\dx\dv,\quad
\Epot(t)=\frac12\int E^2\dx,
\end{equation}
and the total energy $\energy=\energy_{\mathrm{kin},e}+\energy_{\mathrm{kin},i}+\Epot$, driven predominantly through 
the work done by $\Epond$ on the electrons, with a direct contribution to the ions expected to be negligible because 
of the mass-ratio suppression $\mu Z_i\ll1$. \Cref{fig:energies} shows the evolution of these energies for the ion-dynamic run. Adding mobile ions means that the initial total energy is split equally between ions and 
electrons. The total energy increases over the simulation time due to the externally applied ponderomotive force 
peaking at around $t=250\omega_{p,e}^{-1}$, after which the system rapidly loses energy again as the external drive 
is turned off. \Cref{subfig:relative_energy_deviations} and 
\cref{subfig:relative_energy_contribution_only_kinetic_etot_t} confirm that most of the externally introduced energy 
is transferred into the kinetic energy of the electrons. The kinetic energy level of the electrons stabilizes at a level $\sim7$\% higher than the initial one. The contribution to the ion kinetic energy, as well as the energy of the self-induced electric field, is near negligible, which confirms that the applied force is most efficient in inducing 
electron dynamics, while ions play a minor role in the energy cascade for this particular KEEN wave setup.

\begin{figure}
\centering
\begin{subfigure}{0.49\textwidth}
\caption{\label{subfig:relative_energy_contribution}}
\includegraphics[width=0.99\textwidth]{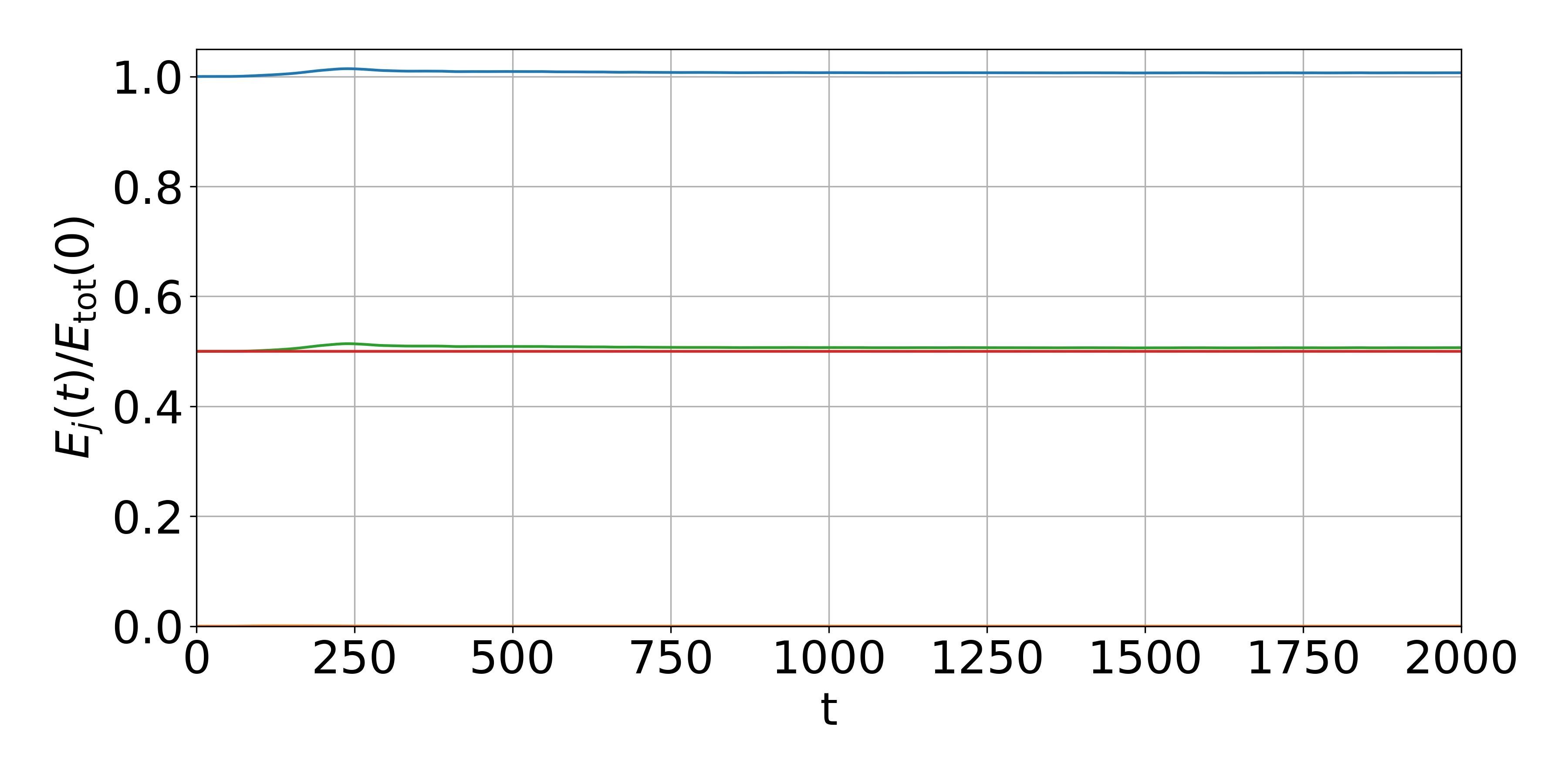}
\end{subfigure}
\begin{subfigure}{0.49\textwidth}
\caption{\label{subfig:relative_energy_deviations}}
\includegraphics[width=0.99\textwidth]{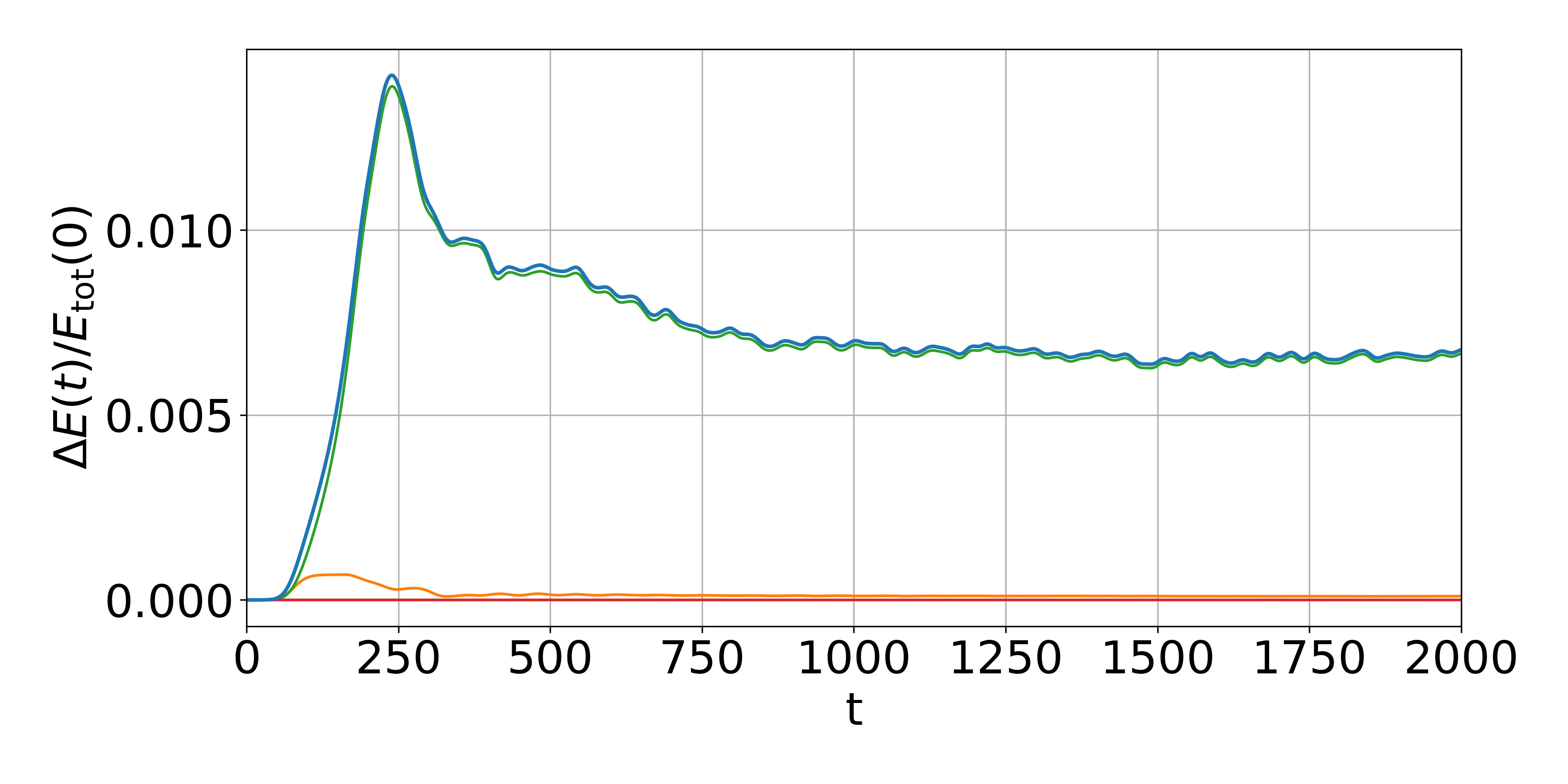}
\end{subfigure}
\begin{subfigure}{0.49\textwidth}
\caption{\label{subfig:relative_energy_contribution_only_kinetic_etot_0}}
\includegraphics[width=0.99\textwidth]{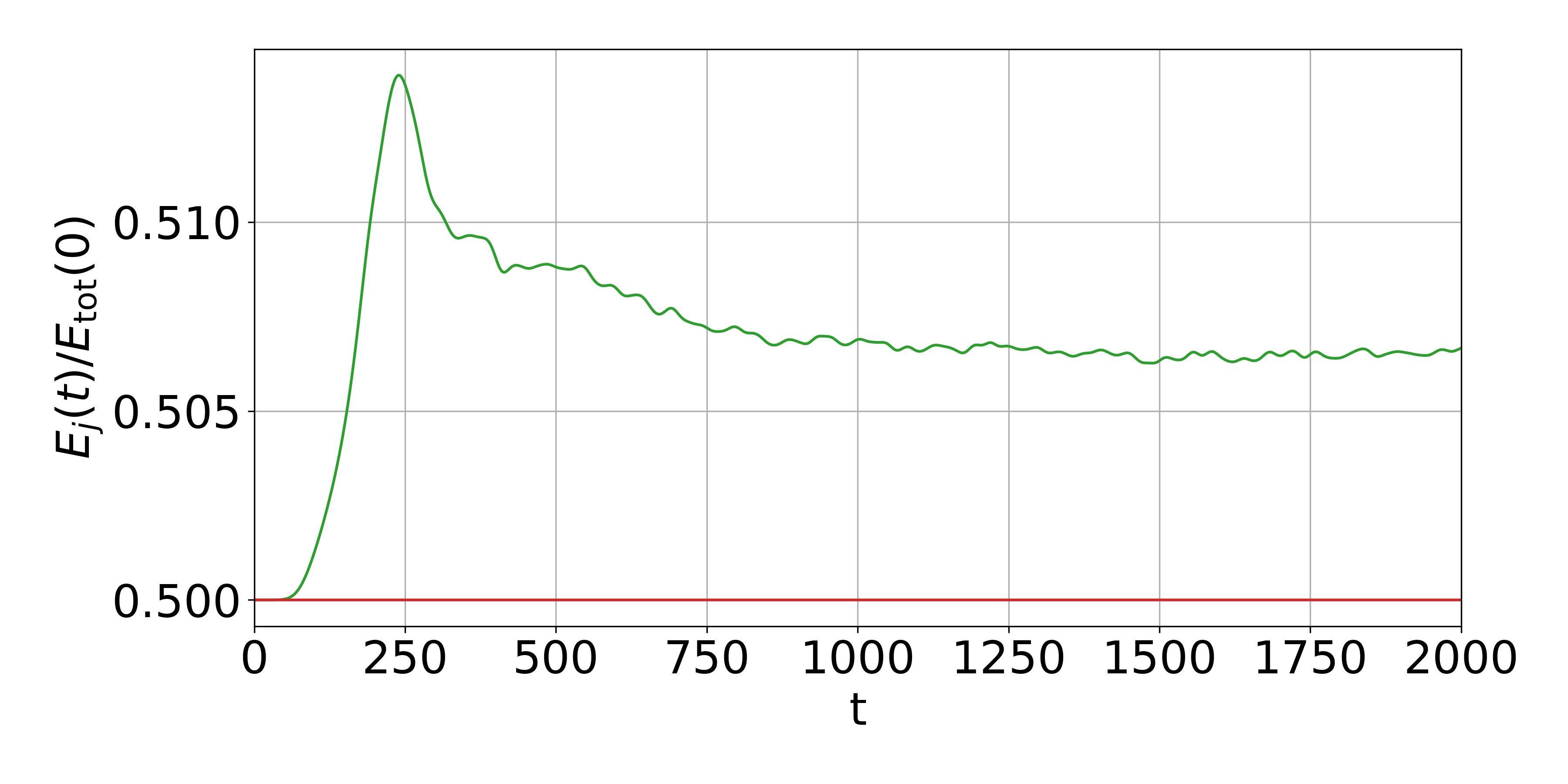}
\end{subfigure}
\begin{subfigure}{0.49\textwidth}
\caption{\label{subfig:relative_energy_contribution_only_kinetic_etot_t}}
\includegraphics[width=0.99\textwidth]{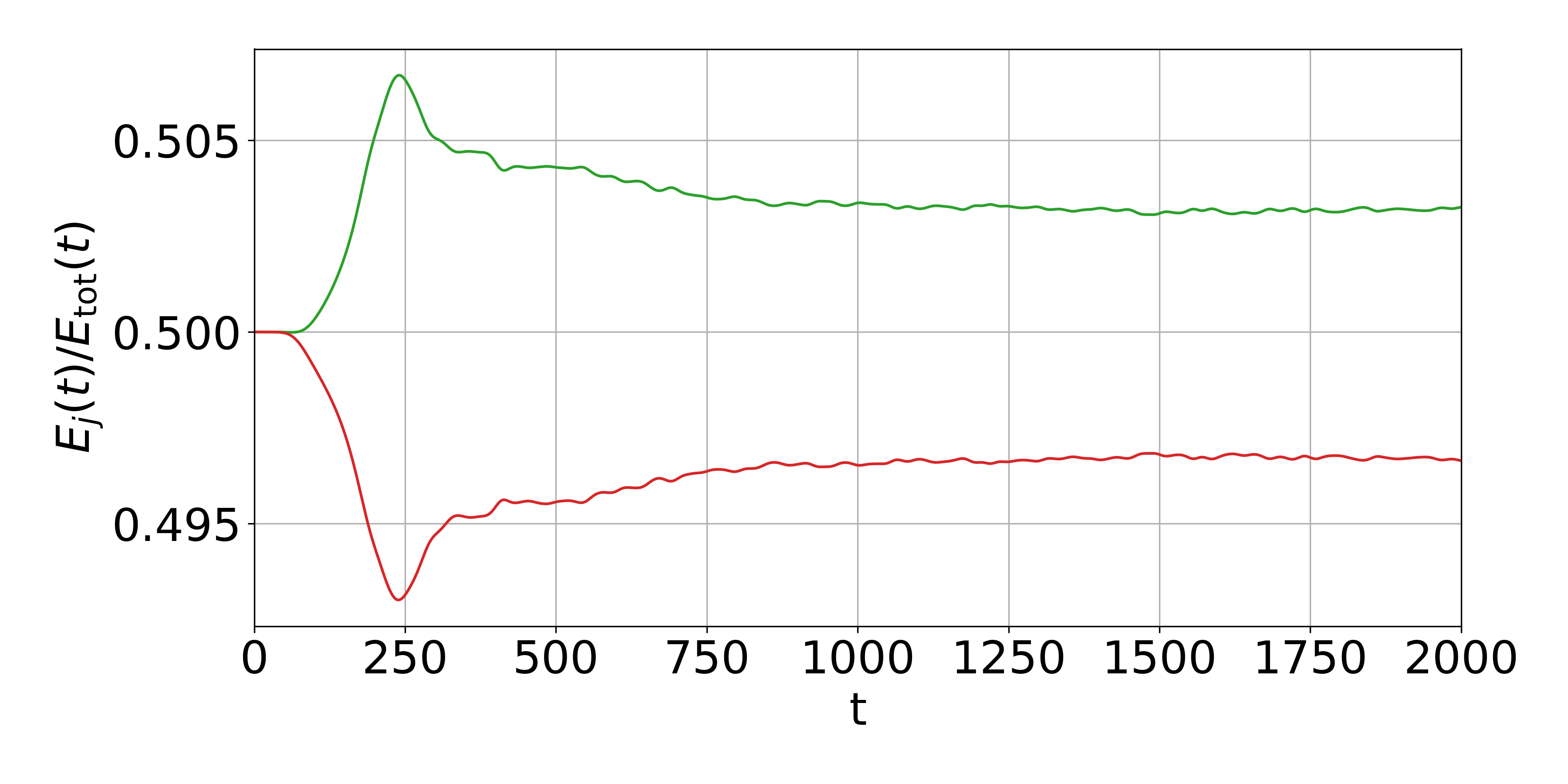}
\end{subfigure}
\begin{subfigure}{0.69\textwidth}
\includegraphics[width=0.99\textwidth]{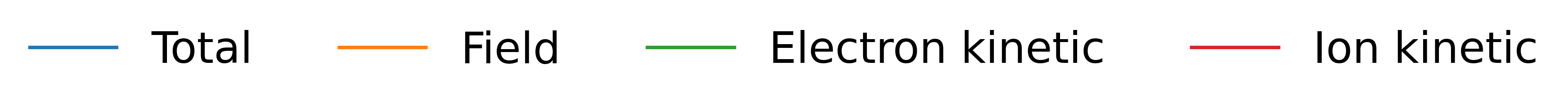}
\end{subfigure}
\caption{Evolution of electron and ion kinetic energy, potential energy and total energy for the ion-dynamic canonical run. In a) we see the relative contribution of each energy component, while b) shows the deviation of each energy component relative to the initial total energy. Further, plot c) highlights the kinetic energy contribution relative to the initial total energy, while plot d) puts it in relation to the current total energy. }
\label{fig:energies}
\end{figure}

\section{Conclusion}

We extended the canonical KEEN wave problem to a two-species Vlasov--Poisson system with dynamical ions, generalizing CMM-\NuFi{} to multiple species by maintaining one independent set of characteristic submaps per species, coupled only through the shared Poisson solve. The resulting multi-map scheme 
resolves the disparate scales imposed by the ponderomotive drive and the large ion-to-electron mass ratio without prohibitively fine grids, making long-time ($t=2000\,\omega_{p,e}^{-1}$), fully kinetic simulations with mobile ions at realistic mass ratios computationally tractable.

Comparing the ion-dynamic run to the ion-static reference, the electron dynamics that seed the KEEN wave are essentially unaffected by ion motion during and shortly after the drive, as expected from the mass-ratio suppression $\mu Z_i\ll1$. At later times, however, the slowly evolving ion response feeds back into the field and produces a clear divergence: additional small-scale electron phase-space vortices, a shifted primary vortex, and a slow drift in the $\rho$-harmonics. The ion distribution stays close to a Maxwellian throughout, and the energy budget confirms that the ponderomotive force drives mainly the electrons, with ion kinetic and field energy contributing only marginally.

Thus, while ion mobility does not qualitatively alter the canonical KEEN wave on these time scales, it is not a passive background and leaves {behind} a measurable, cumulative influence on the electron phase space. 
Beyond this physical picture, the work establishes multi-map CMM-\NuFi{} as a practical tool for kinetic multi-species prob lems{even} with strong scale separation. 
{The compositional adaptivity of the method allows zooming into the fine-scale structure of the solution while using coarse grids of the maps.}
Natural extensions include multiple charged ions with collisions, electromagnetic effects in higher phase-space dimensions, species-dependent remapping and map-grid parameters, and a closer look at the mechanism generating the additional small-scale structures at late times.




%



\subsubsection*{Author Contribution Statement}

The following outlines the authors' contributions to this work. 

{\small
\noindent
\begin{tabular}{@{}p{0.17\linewidth} p{0.77\linewidth}}
    \textbf{R.-P. Wilhelm:} & Numerical Simulations, Implementation, Visualization, Writing original draft\\
    \textbf{P. Krah:} & Initial Idea, Methodology, Investigation, Writing original draft,\\
\textbf{K. Schneider:} & Review \& Editing, \\
\textbf{F. Bacchini:} & Review \& Editing, Supervision, Funding acquisition,\\
\textbf{V. Grandgirard:} & Review \& Editing, Supervision, Funding acquisition.
\end{tabular}
}

\subsubsection*{Code Availability}
The code is publicly available on Github in the NumericalFlowIteration
library: \url{https://github.com/paulwilhelmvlasov/NumericalFlowIteration}. 

\section*{Acknowledgments}
This paper was submitted to Springer Proceedings of the 34th International Symposium on Rarefied Gas Dynamics.

This work was supported by the SPACE CoE, funded by the EU and several partner countries under grant No. 101093441.
F.B.\ acknowledges support from the FED-tWIN programme (profile Prf-2020-004, project ``ENERGY''), issued by BELSPO, and from the FWO Junior Research Project G020224N granted by the Research Foundation -- Flanders (FWO).
{
\bibliographystyle{siamplain}
\bibliography{references}
}


\end{document}